\documentclass[aps,prl,twocolumn,superscriptaddress,nofootinbib]{revtex4-2}
\usepackage{amsmath,amssymb}
\usepackage{graphicx}
\usepackage[colorlinks=true,linkcolor=blue,citecolor=blue,urlcolor=blue]{hyperref}

\newcommand{\jp}{J/\psi}

\begin{document}

\title{When hadrons collide: di-$\jp$  structure in the final state}

\author{Eric S. Swanson}
\email{swansone@pitt.edu}
\affiliation{University of Pittsburgh, Pittsburgh, PA 15260, USA}

\date{\today}

\begin{abstract}
The structures in the di-$\jp$ spectrum seen at the LHC are widely read as fully charmed
tetraquarks. As an alternative scenario, the $2^{++}$ production amplitude for a compact $cc\bar c\bar c$ source
 rescattering through ten charmonium-pair channels is presented.  The model is completely fixed except for three background parameters and provides a competitive description of the data, with $\chi^2/\mathrm{ndf}=56/50$. Prominent features are driven by $\jp\psi(nS)$ thresholds or threshold enhancement, with no need for exotic hadrons.
\end{abstract}

\maketitle

\paragraph*{\textbf{Introduction.}}

In 2020, LHCb found a narrow structure near 6.9~GeV in the $\jp$-pair mass spectrum of $pp\to \jp\jp+X$, together with a broad
enhancement between threshold and 6.8~GeV, a dip near 6.75~GeV, and a hint of a further structure
near 7.2~GeV~\cite{LHCb:2020bwg}. ATLAS~\cite{ATLAS:2023bft} and CMS~\cite{CMS:2023owd} confirmed
the pattern. These structures were subsequently interpreted as resonances: CMS described its spectrum with three interfering Breit--Wigner amplitudes,
$X(6600)$, $X(6900)$ and $X(7100)$. A larger CMS sample establishes all three structures  above $5\sigma$, favours $J^{PC}=2^{++}$ for all of them, and interprets them as radial
excitations of a diquark--antidiquark system~\cite{CMS:2026tiu}.

Most explanations assign the structures to states. Fully charmed tetraquarks were proposed long
ago~\cite{Iwasaki:1975pv,Chao:1980dv,Ader:1981db,Heller:1985cb}, and since the LHCb observation compact
$cc\bar c\bar c$ states have been computed in diquark models~\cite{Bedolla:2019zwg,Giron:2020wpx,Karliner:2020dta}, in constituent
quark models~\cite{Lu:2020cns,Deng:2020iqw} and with QCD sum rules~\cite{Wang:2020ols}, and
perturbative arguments have been given for spin 2~\cite{Zhang:2020hoh}. Coupled-channel analyses
fit unitarised amplitudes for $\jp\jp$, $\jp\psi(2S)$, $\jp\psi(3770)$ and similar channels, with
contact, effective-Lagrangian or Pomeron-exchange interactions, to the data. They also end with
poles that are interpreted as states: a near-threshold $X(6200)$~\cite{Dong:2020nwy,Song:2024ykq}, a
dynamically generated $X(6900)$~\cite{Gong:2020bmg,Liang:2021fzr}, or a compact
$X(6900)$~\cite{Zhou:2022xpd,Kuang:2023vac}. Closest to the present work in method, Ortega
\textit{et al.}\ solve a resonating-group non-confining quark model for charmonium pairs and identify 29 poles
between 6.1 and 7.6~GeV, several of which they assign to the observed
structures~\cite{Ortega:2023pmr}. Combined fits of the three experiments with interfering
resonances have also been made~\cite{Wang:2026kcw}.

A smaller group of papers attributes the structures to thresholds. Wang \textit{et al.}\ reproduce
the LHCb and CMS spectra with charmonium pairs produced directly and converted to $\jp\jp$ through
one-loop cusps, with a fitted coupling and phase for each
channel~\cite{Wang:2020wrp,Wang:2022jmb}; Zhuang \textit{et al.}\ find that $X(6900)$ can be a
$\jp\psi(3770)$ cusp~\cite{Zhuang:2021pci}; and threshold cusps have been discussed in
general~\cite{Dong:2020hxe}.

We will not comment on these studies except to say that it is far too common to invent dynamics that creates a plethora of states and then claim that a small set of these explain the phenomenon in question.

In contrast to all these interpretations, we show that a model whose dynamics is fixed by the charmonium spectrum, combined with a simple background fitted to the di-$\jp$ data, produces the observed pattern by
rescattering alone. In this approach charmonium mesons interact by quark exchange, computed from the same
Hamiltonian that gives the charmonium spectrum. The $2^{++}$ $\jp\jp$ channel couples strongly to
$\jp\psi(nS)$, and each threshold opening removes flux from $\jp\jp$. The resulting dips and the
peaks between them follow from a single $2^{++}$ amplitude, so the interference that a
Breit--Wigner description requires appears automatically.

\paragraph*{\textbf{Constituent Quark Model.}}

The starting point is a reasonably standard nonrelativistic model of charmonia, defined by
\begin{equation}
H=\sum_i\Big(m_i+\frac{p_i^2}{2m_i}\Big)+\sum_{i<j}\mathbf F_i\cdot\mathbf F_j\,V(r_{ij}),
\label{eq:H}
\end{equation}
\begin{equation}
V(r)=\frac{\alpha_s}{r}-\frac{3b}{4}\,r-\frac{8\pi\alpha_s}{3m_im_j}\,
\Big(\frac{\sigma}{\sqrt\pi}\Big)^3e^{-\sigma^2r^2}\,\mathbf S_i\cdot\mathbf S_j ,
\label{eq:V}
\end{equation}
where $\mathbf F$ are colour generators. For a colour-singlet $c\bar c$ pair,
$\mathbf F_i\cdot\mathbf F_j=-4/3$, and $V$ becomes the Coulomb, linear and smeared contact
hyperfine potential of Ref.~\cite{Barnes:2005pb}, whose parameters, $m_c=1.479$~GeV,
$\alpha_s=0.546$, $b=0.143$~GeV$^2$ and $\sigma=1.095$~GeV, are fixed by the charmonium spectrum.
We neglect spin-orbit and tensor interactions because they make tiny contributions to the
scattering considered here. To facilitate the subsequent scattering computations 
meson wavefunctions are expanded in a Gaussian basis~\cite{Swanson:1992ec}.

\paragraph*{\textbf{Production.}}
We assume that two gluons make a $cc\bar c\bar c$ system through $g\to c\bar c$ splitting. We take the source
to be compact, in colour $|8_{13}8_{24}\rangle$ and spin $|(11)_2\rangle$, and model the spatial distribution as:
\begin{equation}
\Phi_\beta=e^{-\beta(r_{13}^2/2+r_{24}^2/2+R^2)}\,|8_{13}8_{24}\rangle\,|(11)_2\rangle ,
\end{equation}
with $\beta$ of order $m_c^2$. Because $\Phi_\beta$ is symmetric in all quark coordinates
($r_{13}^2/2+r_{24}^2/2+R^2=\sum_i(\mathbf x_i-\mathbf X)^2$ for equal masses) and has $S=2$, the
antisymmetriser projects it onto the diquark--antidiquark colour state $|\bar3_{12}3_{34}\rangle$. A
colour-singlet admixture $|1_{13}1_{24}\rangle$, which leading-order $gg\to cc\bar c\bar c$ also
produces, therefore only changes the normalisation and leaves the shape of the spectrum unchanged.

The rate into final channel $a$ is
\begin{equation}
\frac{dN_a}{dM}\propto|F_a(M)|^2,\qquad F_a(M)=\big\langle\Psi^{(-)}_a(M)\big|\mathcal A\,\Phi_\beta\big\rangle ,
\label{eq:F}
\end{equation}
where $\Psi^{(-)}_a$ is the coupled-channel scattering solution with incoming-wave boundary
conditions in channel $a$, normalised to unit flux. It is therefore energy-normalised (up to a
factor $2\pi$), and phase space is included. This is the standard final-state-interaction formula~\cite{Watson:1952ji}.
$\Psi^{(-)}_a$ contains every channel $\gamma$, so $F_a$ sums direct production into $a$ and
production into each $\gamma$ followed by the transition $\gamma\to a$, to all orders in quark
exchange. Below the first inelastic threshold the phase of $F_a$ is the elastic phase shift.
Equivalently, $F$ has the familiar P-vector form $F\propto(1-iK)^{-1}P$ in the open channels, with
$K$ the model's K-matrix. We have checked numerically that the $P$ implied by Eq.~\eqref{eq:F} is real,
to within the unitarity error of the solution. Here, however, $P$ is computed from the source and
the quark model rather than fitted. Since $\Phi_\beta$ has no overlap with the singlet pairing
$(13)(24)$, the $\jp\jp$ yield comes from the $(14)(23)$ component, with colour weight $\sqrt8/3$.
We characterise the distortion by the enhancement
\begin{equation}
E(M)=\frac{|F_{\jp\jp}(M)|^2}{|F^{(0)}_{\jp\jp}(M)|^2},
\end{equation}
where $F^{(0)}$ is computed with the inter-meson kernel switched off. $E$ is a final-state
interaction factor; phase space and the parton luminosity cancel in it. As confirmed by direct computation, its shape is approximately independent of
the source size for $\beta\ge0.5$~GeV$^2$ (rms $c\bar c$ separation below 0.34~fm) and saturates by
$\beta\approx16$~GeV$^2$. We use $\beta=4$~GeV$^2$ in the following.

\paragraph*{\textbf{Rescattering.}}
The direct terms of Eq.~\eqref{eq:H} vanish
between two colour-singlet mesons 
because each
meson's colour generators sum to zero. Mesons thus interact through quark interchange: an
interaction between constituents of different mesons accompanied by the exchange of a quark or
antiquark between them. Barnes and Swanson evaluated these processes as quark Born
diagrams~\cite{Barnes:1991em,Swanson:1992ec}, and the same formalism has described $I=2$ $\pi\pi$ and
other exotic meson--meson channels~\cite{Barnes:1991em,Barnes:2000hu}, kaon--nucleon
scattering~\cite{Barnes:1992ca}, $B\bar B$ intermeson potentials~\cite{Barnes:1999hs},
$\jp$ and $\psi'$ dissociation by light hadrons~\cite{Wong:1999zb,Wong:2001td}, and the
short-range structure of the $X(3872)$~\cite{Swanson:2003tb}. Here we go beyond Born order. The
exchange kernels $\langle\phi_A\phi_B|(H-E)P_{ij}|\phi_C\phi_D\rangle$, together with the overlap kernel
of the antisymmetrised states, appear in coupled-channel resonating-group
equations~\cite{Wheeler:1937zz}, which contain quark exchange to all orders. The 
 variational resonating-group method in the form of  Kohn--Hulth\'en--Kato~\cite{Kohn:1948col,Hulthen:1948,Kato:1951,Kamimura:1977okl} is then used to obtain scattering solutions. For trial solutions $\Psi^{(\alpha)}_{(t)}$ with unit incoming
flux in channel $\alpha$ and outgoing amplitudes $S_{\beta\alpha(t)}$, the bilinear functional
\begin{equation}
S_{\beta\alpha}[\Psi_{(t)}]=S_{\beta\alpha(t)}+i\int\tilde\Psi^{(\beta)}_{(t)}\,(H-E)\,\Psi^{(\alpha)}_{(t)}\,d\tau
\end{equation}
($d\tau$ denotes a sum over colour and spin indices and an integral over the Jacobi coordinates)
is stationary, $\delta S_{\beta\alpha}=0$, at the exact solution. Its stationary value gives the
S-matrix with an error of second order in that of the trial functions. Here
$\tilde\Psi^{(\beta)}$ is the time-reversed solution. The relative wavefunctions are expanded in
Gaussians inside a matching radius $R_c$ and matched to Hankel functions outside it. $S$ is  
symmetric by construction. Unitarity is not imposed; we have confirmed that it holds to better than $10^{-3}$ in all calculations reported here.
Finally, the nonorthogonal Gaussian basis can be ill-conditioned, so we use a truncated singular-value solution.

In view of the CMS spin analysis~\cite{CMS:2026tiu}, we work in the $J^{PC}=2^{++}$ sector with total quark spin $S=2$ and relative S-waves. The ten channels considered here are $\jp\jp$, $\jp\psi(2S)$,
$\chi_{c0}\chi_{c2}$, $\chi_{c1}\chi_{c1}$, $h_ch_c$, $\chi_{c1}\chi_{c2}$, $\chi_{c2}\chi_{c2}$,
$\jp\psi(3S)$, $\psi(2S)\psi(2S)$ and $\jp\psi(4S)$. P-wave pairs enter through their recoupled
$(L,S)$ components. The $\chi_{cJ}\chi_{cJ'}$ pairs contain both the $(L,S)=(0,2)$ component of
S-wave $\jp\jp$ and the $(2,0)$ component of $h_ch_c$, which is how $h_ch_c$ couples. Model
thresholds lie within 30~MeV of the physical ones. Adding D-wave channels that couple to the  $\chi_{cJ}$
components has little effect, and the S- to D-wave $\jp\jp$ amplitude
stays below 0.002.

\begin{figure}[t]
\includegraphics[width=\columnwidth]{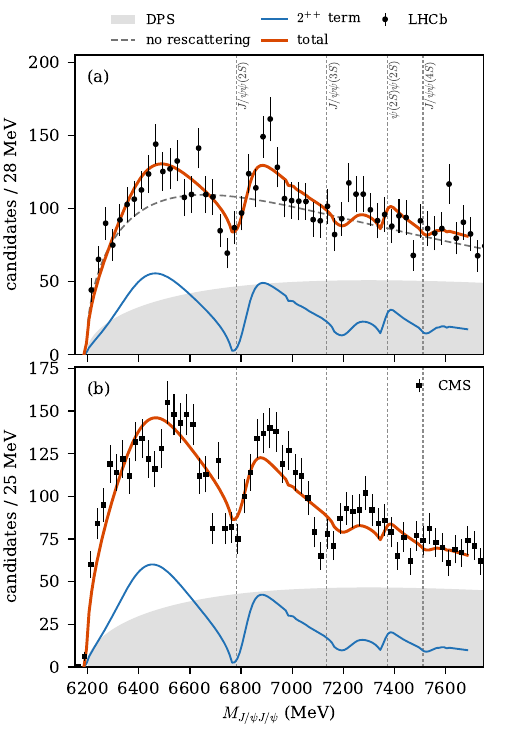}
\caption{(a) LHCb di-$\jp$ spectrum~\cite{LHCb:2020bwg} and Eq.~\eqref{eq:fit} with $E$ from the
ten-channel calculation at model masses: total (solid), the $2^{++}$ term $a\,E\,N_{\rm SPS}$ (blue),
continuum without rescattering (dashed) and DPS (shaded). Vertical lines mark the physical
thresholds. (b) Independent fit to the CMS spectrum~\cite{CMS:2023owd}.}
\label{fig:fit}
\end{figure}

The rescattering contribution, whose shape is fixed by the model, is shown as the blue curves in Fig.~\ref{fig:fit}.
One observes a broad $\jp\jp$ enhancement near 6.5~GeV, in the $X(6600)$ region. This is
a threshold enhancement from the elastic $\jp\jp$ interaction, sharpened on its upper side by the
dip that follows. This dip, near 6800 MeV, is due to flux entering the $\jp\psi(2S)$ channel. This, in turn, leads to a peak near 6900 MeV that can be identified with the putative $X(6900)$. Structure at higher energy is due to coupling to additional channels and to dips caused by the $\jp\psi(nS)$ thresholds.
Every feature is tied to a
threshold with positions fixed by the charmonium spectrum and the subsequent  quark-exchange
couplings -- no parameters have been adjusted.

\paragraph*{\textbf{Comparison with data.}}
To compare with the measured spectra we need the yield without rescattering, which depends on
parton luminosities and acceptance. We therefore draw on LHCb's 
fit, using its single-parton (NRSPS) and double-parton (DPS)
templates~\cite{Kom:2011bd,Lansberg:2014swa}, and let the model supply only the final-state
distortion. This gives an expression for the differential rate of the form
\begin{equation}
\frac{dN}{dM}=\big[a\,E(M)+c\big]\,N_{\rm SPS}(M)\,e^{-\lambda x}+b\,N_{\rm DPS}(M),
\label{eq:fit}
\end{equation}
with $x=M-6.2$~GeV. The term $a\,E$ contains the full interference between the unscattered and the
rescattered $2^{++}$ amplitude; the dips are this interference. The term $c$ collects
single-parton production in other partial waves, which does not interfere with it after angular
integration. Double-parton scattering produces the two $\jp$ in independent collisions, so it neither
rescatters nor interferes and adds to the rate; it is fixed at LHCb's yield ($b=1$). Finally,
$\lambda$ corrects LHCb's continuum slope. We use the model masses directly. Model thresholds
differ from the physical ones by up to 28~MeV, so the features may appear shifted by similar
amounts; mapping the model thresholds onto the physical ones changes $\chi^2$ little (61/50).
We only fit the range 6.2--7.7~GeV, which can be reasonably covered by the ten channels considered.

Figure~\ref{fig:fit}(a) shows the result. Three parameters, $a=1.00$, $c=1.29$ and
$\lambda=0.75$~GeV$^{-1}$, give $\chi^2/\mathrm{ndf}=56/50$. The ratio $a/(a+c)=0.44$ is the fraction
of single-parton production in the $2^{++}$ S-wave. If all of it is taken to rescatter ($c=0$),
the fit gives $\chi^2/\mathrm{ndf}=90/51$: evidently the data require the incoherent remainder. With LHCb's
continuum slope ($\lambda=0$) the fit gives $152/51$. The fit reproduces the broad enhancement
above threshold, the dip near the $\jp\psi(2S)$ threshold, the peak at 6.9~GeV and the shoulder
near 7.2~GeV between the $\jp\psi(3S)$ and $\psi(2S)\psi(2S)$ thresholds. Adding a $0^{++}$
enhancement term computed in the same way improves $\chi^2$ by only 1--3.

CMS [Fig.~\ref{fig:fit}(b)] gives $\chi^2/\mathrm{ndf}=117/55$ with its own $a$, $b$, $c$ and
$\lambda$. The model reproduces the overall pattern again, but CMS's dips are deeper than the
model's. The model also undershoots
the 6.9~GeV peak in both data sets (by 15--20\% at the maximum for LHCb) and has a small $\psi(2S)\psi(2S)$
threshold peak near 7.4~GeV that the data do not show.

Overall, given that the model is completely specified by a quark model representation of the low-lying charmonium spectrum (except for 3 or 4 background parameters), the general agreement with data is striking.

\paragraph*{\textbf{Poles.}}
The amplitude has poles, and it is natural to ask whether they are the reported states. In the
coupled-channel system the main ones lie at $6809-68i$~MeV, on the sheet adjacent to the physical
region above the $\jp\psi(2S)$ threshold, and at $7102-57i$~MeV, above the $\chi_{c1}\chi_{c2}$
threshold. Shifted to the physical thresholds, they lie near 6830 and 7107~MeV, in the region of
$X(6900)$ and $X(7100)$, although the second couples only weakly to $\jp\jp$.

We stress, however, that these poles are not states in the usual sense. Each comes from an interaction that is repulsive
at low momentum and attractive above a crossover momentum, the equivalent of a repulsive shell
around an attractive core. This crossover occurs because of competition between the four interaction terms in the exchange mechanism~\cite{Swanson:1992ec} and has little to do with quark flavour or resonances. Thus the poles are simply reflections of the form of a quantum mechanical interaction in the complex plane, and their presence should not be used to infer the existence of putative tetraquarks. For this reason, we resist an overinterpretation of the amplitude poles and focus attention on the observable and its description.

\paragraph*{\textbf{Discussion.}}
The rescattering picture makes definite predictions. For example, the same thresholds should appear in the
$\jp\psi(2S)$ spectrum, where the $\jp\psi(2S)$ channel is now the final state and the
$\chi_{cJ}\chi_{cJ'}$ and higher thresholds should cause structure. ATLAS~\cite{ATLAS:2023bft} and
CMS~\cite{CMS:2026tiu} see structure in this spectrum, which CMS finds consistent with $X(6900)$ and
$X(7100)$.

\begin{figure}[t]
\includegraphics[width=\columnwidth]{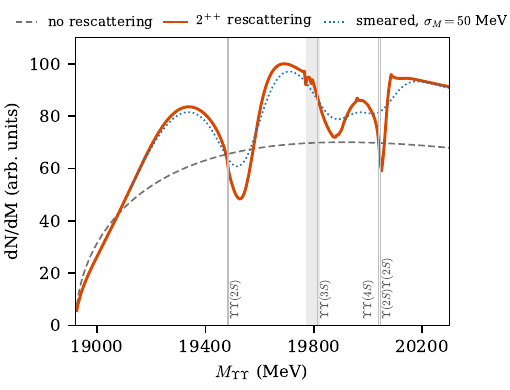}
\caption{Predicted $\Upsilon(1S)\Upsilon(1S)$ spectrum (arbitrary normalisation): continuum
without rescattering (dashed), with $2^{++}$ rescattering through ten bottomonium-pair channels
(solid), and the same smeared with a 50~MeV resolution (dotted). Thin vertical lines mark the
physical thresholds; the shaded band covers the $\chi_{bJ}\chi_{bJ'}$ and $h_bh_b$ thresholds.}
\label{fig:upsilon}
\end{figure}

The mechanism also applies to bottomonium pairs, for which we make a prediction. We refit $m_b$,
$\alpha_s$ and $\sigma$ in Eq.~\eqref{eq:V} to thirteen bottomonium levels below $B\bar B$
threshold and keep the string tension at its charmonium value. This gives $m_b=4.843$~GeV,
$\alpha_s=0.426$ and $\sigma=2.04$~GeV, with an rms deviation of 21~MeV. The ten channels are the
analogues of those above, with $\Upsilon(nS)$, $\chi_{bJ}$ and $h_b$ in place of $\psi(nS)$,
$\chi_{cJ}$ and $h_c$. The channel thresholds are set to the physical meson masses, while the
kernels use the model wavefunctions and energies. Because $\Upsilon\Upsilon$ momenta reach 3.5~GeV, we use a matching radius of $R_c=6$~GeV$^{-1}$ and scale the source size with the heavy-quark mass, $\beta=4\,(m_b/m_c)^2\approx43$~GeV$^2$, to account
for the smaller bottomonium states. The continuum is taken as
$\rho_\Upsilon(M)\,e^{-c_\Upsilon(M-2m_\Upsilon)}$, with $\rho_\Upsilon$ the two-body phase-space
factor. The slope is scaled from charm by the inverse mass ratio, $c_\Upsilon\approx0.47$~GeV$^{-1}$,
as for a parton luminosity falling as a power of $M$, and $a/c$ is taken from the LHCb fit. DPS is
omitted; it would add a smooth background and dilute the structure. The predicted spectrum
(Fig.~\ref{fig:upsilon}) shows broad enhancements near 19.3 and 19.7 GeV, a dip near 19.5 GeV driven by the $\Upsilon\Upsilon(2S)$ coupling, a broad dip near 19.9 GeV driven by $\Upsilon\Upsilon(3S)$, and a notably sharp dip near 20.05 GeV due to coupling to the nearly degenerate $\Upsilon\Upsilon(4S)$ and $\Upsilon(2S)\Upsilon(2S)$ channels.

We note that  Wang \textit{et al.}\ predicted
enhancements near 19.3 and 19.7~GeV from their loop mechanism~\cite{Wang:2020wrp}, so the
$\Upsilon\Upsilon$ spectrum tests threshold rescattering in general rather than distinguishing the two
calculations. $\Upsilon(1S)$ pairs have been observed by CMS~\cite{CMS:2016liw,CMS:2020qwa}, and LHCb
has searched for $b b\bar b\bar b$ states~\cite{LHCb:2018uwm}. These searches looked for narrow
resonances and would not be sensitive to the broad features predicted here.

Several physical effects are missing from the calculation. (i) More channels can always be considered, although we expect that the important ones up to 7600 MeV have been accounted for. It is possible that increasingly broad resonances can smear the effects out higher in the spectrum. Several analyses model the $X(6900)$ with $\jp\,\psi(3770)$ content. This channel couples to S-wave $\jp\jp$ with $S=2$ only through tensor interactions, and is therefore ignored in the present model. (ii) Flip-flop or flux tube models of quark dynamics may affect details but should not alter the main features driven by channel coupling. (iii) The source is a single Gaussian with octet--octet colour. The transverse-momentum and rapidity dependence of production is not
      modelled. Furthermore, the  continuum is taken from LHCb's templates with an adjusted slope. It would also be desirable to explain the relative weight between the $J^{PC}=0^{++}$ and $2^{++}$ spectra. (iv) Perhaps the most important missing component is quark annihilation. At the level of mixing in $cc\bar c\bar c$ this is a short range effect that can be expected to have little influence. However, it also permits reactions such as $\jp\jp \to D^{(*)}\bar  D^{(*)}$, which lie far below the region of interest and act as absorption.

These effects might be expected to change the shape of the emergent structures, particularly above 7.1~GeV,
rather than their origin: the structures are tied to charmonium-pair thresholds, and their sizes
come from fixed quark-exchange couplings.

In summary, the di-$\jp$ structures can be understood as rescattering of $2^{++}$ charmonium pairs
produced by a compact source, with no tetraquark resonances required.

\paragraph*{\textbf{AI Statement.}} Claude Opus 5.5 was used for grammar and spell checking, for image generation, to convert an ancient VRGM code to modern C++, and assistance in generalizing that code to coupled channels.

\begin{acknowledgments}
The author acknowledges support from the U.S.\ Department of Energy under Grant No.\ DE-SC0019232.
\end{acknowledgments}

\bibliography{refs}

\end{document}